\documentclass{PoS}

\newcommand{\pks}{PKS~0447$-$439 }

\title{Discovery of VHE emission from \pks with H.E.S.S. and MWL studies}

\ShortTitle{A. Zech et al., \pks in VHE $\gamma$-rays}

\author{A.~Zech$^{1*}$, B.~Behera$^{2\otimes}$,  Y.~Becherini$^3$, C.~Boisson$^1$, B.~Giebels$^4$, M.~Hauser$^2$, M.~Kastendieck$^5$, S.~Kaufmann$^2$, K.~Kosack$^6$, J.-P.~Lenain$^1$, M.~de Naurois$^4$, M.~Punch$^3$, M.~Raue$^5$, H.~Sol$^1$, S.~Wagner$^2$ and the H.E.S.S. collaboration\\ 
\llap{$^1$}Laboratoire Univers et Th\'eories (LUTH), Observatoire de Paris, CNRS, Universit\'e Paris 7 Denis Diderot, Meudon, France\\
\llap{$^2$}Landessternwarte, Universit\"at Heidelberg, Heidelberg, Germany\\
\llap{$^3$}Astroparticule et Cosmologie (APC), CNRS, Universite Paris 7 Denis Diderot, Paris, France\\
\llap{$^4$}Laboratoire Leprince-Ringuet, Ecole Polytechnique, CNRS/IN2P3, Palaiseau, France\\
\llap{$^5$}Institut f\"ur Experimentalphysik, Universit\"at Hamburg, Hamburg, Germany\\
\llap{$^6$}Institut de Recherche sur les lois Fondamentales de l'Univers (IRFU), CEA Saclay, Saclay France\\
E-mail:\email{ $^{*}$Andreas.Zech@obspm.fr},\email{ $^{\otimes}$bbehera@lsw.uni-heidelberg.de}}

\abstract{Very-high energy (VHE) emission has been detected from PKS~0447-439 with the H.E.S.S.
Cherenkov telescope array. This blazar is one of the brightest hard-spectrum extragalactic
objects in the Fermi bright source list. Its detection with H.E.S.S. triggered Target of Opportunity
observations with the {\it Swift} and {\it RXTE} telescopes, which show rapid flaring in the X-ray
band.
The spectrum and light curve measured by H.E.S.S. are presented. Along with the {\it Fermi} LAT
data it is possible to put an upper limit on the redshift of the source. Implications of the 
flux evolution are discussed briefly. }

\FullConference{25th Texas Symposium on Relativistic Astrophysics - TEXAS 2010\\
		December 06-10, 2010\\
		Heidelberg, Germany}

\begin{document}

\section{Discovery of VHE $\gamma$-rays from \pks}

The BL Lac object \pks is one of the brightest blazars in the Fermi Bright Source List \cite{abd2009} and is thought to be at high redshift. A total of 13.5 hours (live time) of good quality H.E.S.S. (High Energy Stereoscopic System) data were taken between November 2009 and January 2010. Data were analysed using the \textit{model} analysis \cite{den2009}, which yields more than two times higher significance for this source than the standard Hillas-type analyses. The data yields a strong VHE (very high energy, $\gtrsim$ 100 GeV) signal at a statistical significance of 13.8$\sigma$ \cite{rau2009}. The flux measured in 
December 2009 above a threshold energy of 250 GeV was $\sim$ 4.5$\%$ of the Crab Nebula flux (from H.E.S.S measurements above the same threshold). The best fit position of the VHE $
\gamma$-ray excess (\mbox{$\alpha_{J2000}=4^\mathrm{h}49^\mathrm{m}29.9^\mathrm{s}$}, \mbox{$\delta_{J2000}=-43^{\circ}50\mathrm{'} 11\mathrm{''}$}) is in 
agreement with the nominal position of the source (\mbox{$\alpha_{J2000}=4^\mathrm{h}49^\mathrm{m}24.7^\mathrm{s}$}, \mbox{$\delta_{J2000}=-43^{\circ}50\mathrm{'} 9\mathrm{''}$}). 
The H.E.S.S. measurements were cross-checked with independent analysis procedures and calibration chains, giving consistent results.

\begin{figure}[!h]
\begin{center}
  \includegraphics[width=.6\textwidth]{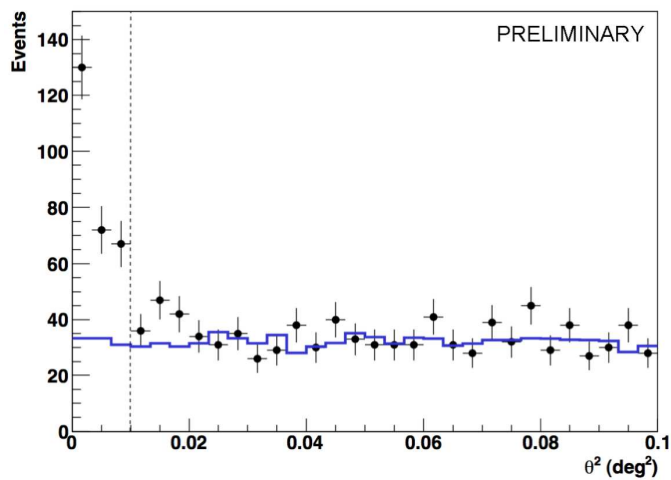}
  \end{center}
  \caption{Theta Square distribution of the $\gamma$-ray excess. Points indicate on-source events, while the off-source event distribution is described by a solid line.  The vertical dashed line indicates the on-source integration region. }
  \label{fig:theta}
\end{figure}

Fig.~\ref{fig:theta} shows the distribution of the squared angular distance of the $\gamma$-ray excess from the nominal source position for on-source events and normalized off-source events. The event distribution is consistent with a point source.

\section{H.E.S.S. spectrum and light curve}

\begin{figure}
\begin{center}
 \includegraphics[width=.6\textwidth]{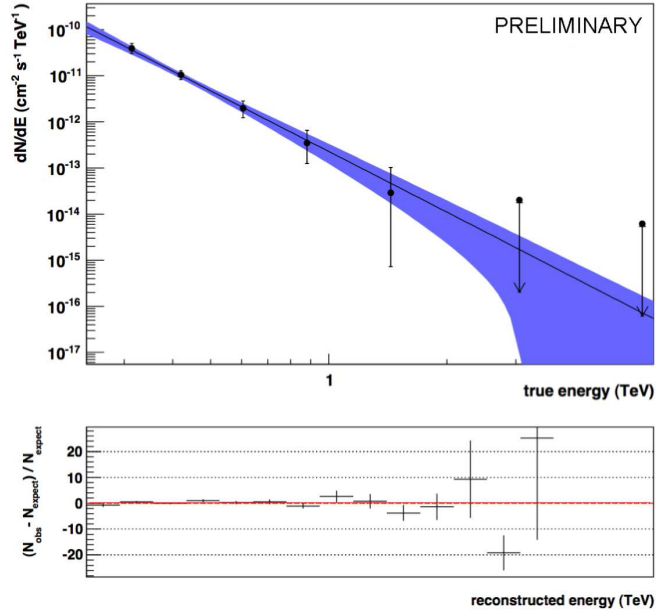}
 \end{center}
  \caption{H.E.S.S. spectrum of the source. The differential flux points and the 68\% confidence band are shown in the upper panel. Residuals of the maximum likelihood fit
  are given in the lower panel.}
   \label{fig:spec}
\end{figure}

The spectrum measured by H.E.S.S. is shown in Fig.~\ref{fig:spec}. It has been extracted using the forward folding method. The blue band indicates the 68$\%$ confidence limit of the best fit power law model ($\chi^2/dof=21.05/16$). Data points with statistical error bars have been calculated from the residuals and have been added to the spectrum plot. The derived VHE spectrum is very soft, with a photon index, $\Gamma =$ 4.36 $\pm$ 0.49 and a flux normalisation, \mbox{$\Phi_{1 \mathrm{TeV}}=(0.23\pm0.11)\times10^{-12}\mathrm{cm}^{-2}\mathrm{s}^{-1}\mathrm{TeV}^{-1}$}. There is no indication for a spectral break or curvature in the observed spectrum.

\begin{figure}[h!]
\begin{center}
   \includegraphics[width=.6\textwidth]{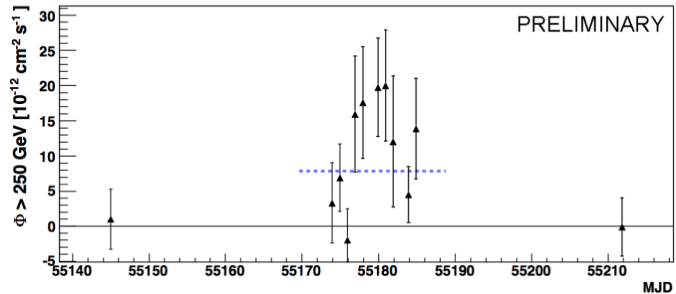}
   \end{center}
  \caption{H.E.S.S. light curve  from November 2009 to January 2010 with a constant fit to the December data (dotted line).}
  \label{fig:lightcurve_hess}
\end{figure}

The nightly binned integrated flux above 250 GeV measured with H.E.S.S. is shown in Fig.~\ref{fig:lightcurve_hess}. Most of the data was taken in December 2009 
(12.17 hours live time of good quality data). 
For November 2009 and January 2010 only 0.86 hours and 0.44 hours of good quality data were available, respectively, and no significant signal was detected. 

A fit of a constant flux to the nightly binned fluxes in December 2009 yields an average flux level of \mbox{(7.9$\pm$1.9)$\times$10$^{-12}$ cm$^{-2}$ s$^{-1}$}  with a $\chi^2$ of 14.87 for 9 d.o.f. (chance probability of 0.09). When including the November and January nights in the fit, a lower average flux of $(5.78\pm1.59) \times10^{-12}$ cm$^{-2}$ s$^{-1}$ with a $\chi^2$ of 19.39 for 11 d.o.f. (chance probability of 0.05) is found. This is a marginal indication for a higher flux in December 2009 compared to the previous and following month, but given that the statistics for these months are very limited, no claim of variability can be made based on the available data.

\section{Multi-wavelength data}
\label{sec:mwl_lc}

\begin{figure}
\begin{center}
 \includegraphics[width=.7\textwidth]{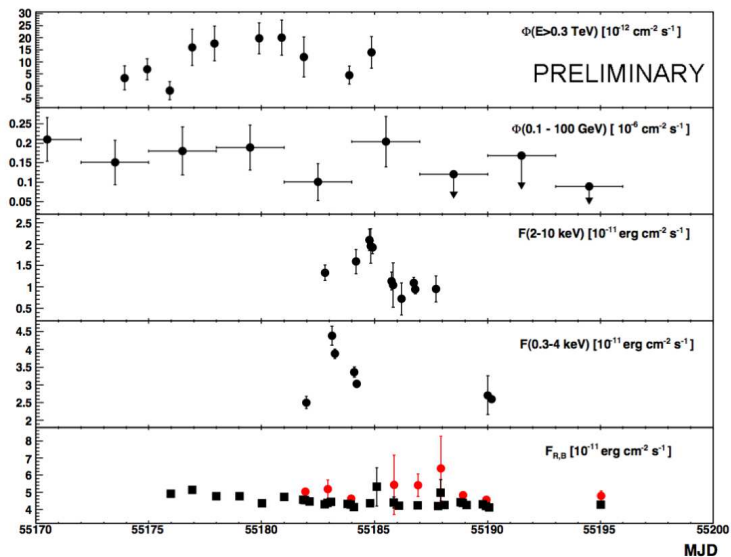}
 \end{center}
  \caption{Multi-wavelength light curves from December 2009. The data points are from H.E.S.S. (first panel), {\it Fermi} LAT (second panel), {\it RXTE} (third panel), {\it Swift} XRT (fourth panel) and ATOM (last panel, R-band: black squares, B-band: red circles ).}
  \label{fig:lc_mwl}
\end{figure}

Fig.~\ref{fig:lc_mwl} shows a comparison of the flux evolution in different energy bands during the month of December 2009.
During the H.E.S.S. observations, the {\it Fermi} LAT light curve shows no significant deviation from a constant flux. 
In the hard X-ray band (2-10 keV), data from the {\it RXTE} show a flare with flux variation of about a factor of 2 on MJD 55185 (December 20th).
A flare on MJD 55183 (December 18th) is seen in the soft X-ray band (0.3-4 keV) in the {\it Swift} XRT data.
It should be noted that none of the pointings of the two X-ray telescopes were exactly simultaneous, with some overlap occurring only during one pointing on MJD 55184 (December 19th). 
The X-ray data have been corrected for Galactic absorption assuming a hydrogen column density n$_H =$ 1.78 $\times$ 10$^{20}$ cm$^{-2}$ \cite{dic1990}.
In the optical band, the ATOM light curve shows a flux decrease in the R band in December 2009. The optical flux has been corrected for Galactic absorption using extinction coefficients
from \cite{sch1998}.
Additional long-term optical data from the ROTSE telescope (not shown here) indicate that the optical flux was in a relatively high state during the H.E.S.S. observations.

\section{Upper Limit on the redshift}

The redshift of \pks is not well known. A value of z$\approx$0.205 was claimed by \cite{per1998}, but is based on very weak spectral features and has not been independently confirmed. 
The most recent study by \cite{lan2008} provides only a lower limit of 0.176 based on photometric estimates.\footnote{It should be noted that the value of z=0.107 provided by the SIMBAD
database and based on \cite{cra1997} is incorrect (private communication with H.~Landt and M.~V\'eron-Cetty).}

\begin{figure}[!h]
\begin{center}
 \includegraphics[width=.65\textwidth]{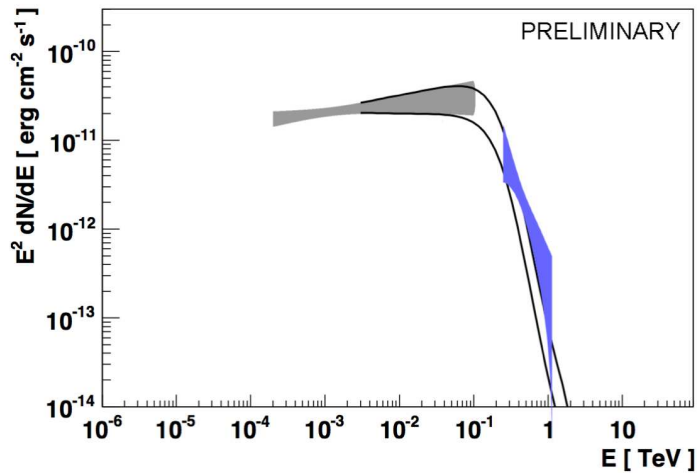}
 \end{center}
  \caption{Extrapolated {\it Fermi} LAT spectrum (grey bowtie, black lines), absorbed by the EBL assuming z=0.53, in comparison with the measured (uncorrected) H.E.S.S. spectrum (blue bowtie). The {\it Fermi} LAT and H.E.S.S. bowties indicate the 2$\sigma$ confidence band, assuming Gaussian errors.}
  \label{fig:zlimit}
\end{figure}

The combined {\it Fermi} LAT and H.E.S.S. data set is used to derive an upper limit on the redshift of the source with no assumptions on the source emission characteristics, other than excluding spectral upturns beyond the {\it Fermi} LAT energy band, which are not expected and would be theoretically difficult to account for. The {\it Fermi} LAT data from MJD 55050 to MJD 55180 have been extrapolated up to VHE energies and corrected for EBL absorption using the model from \cite{fra2008}. Fig. \ref{fig:zlimit} shows the result for z=0.53.  

The two black lines indicate the extrapolated and absorbed {\it Fermi} LAT spectrum corresponding to the upper and lower limit of the {\it Fermi} LAT bowtie. At the chosen redshift, the measured H.E.S.S. spectrum is significantly harder than the spectrum expected from an extrapolation of the {\it Fermi} LAT data. An upper limit of about z<0.53 can be put on the redshift of the source at the 95$\%$ confidence level. Intrinsic softening of the spectrum beyond the {\it Fermi} band, which might occur in this source, would only lead to a lower value for the inferred redshift.

If there was no intrinsic change in the spectrum in the high energy and VHE energy range, a redshift of z$\approx$0.42 would provide the closest agreement of the extrapolated and absorbed
{\it Fermi} LAT flux with the measured H.E.S.S. flux.

\section{Conclusions}

The rapid flaring in the X-ray range, of the time scale of a day,  is not reflected in the {\it Fermi} LAT and VHE energy bands, but might well be hidden in the statistical uncertainties of those data. 
No rapid variability is detected in the optical band either. 

Estimations based on a comparison of the extrapolated {\it Fermi LAT} spectrum with the measured H.E.S.S. spectrum indicate a relatively high redshift of the source, if no intrinsic spectral
changes are present. An upper limit on the redshift has been derived at about z<0.53. 

In the SSC framework, the flux variability in the X-ray data of about one day constrains the ratio of the size of the emission region $R$ over the Doppler factor $\delta$ to $R\delta^{-1} \le$
2.2$\times$10$^{15}$cm for the lower redshift limit of z=0.176 and to $R\delta^{-1} \le$1.7$\times$10$^{15}$cm for the upper limit of z=0.53. These values are not unusual for BL Lac objects.

A detailed study of the upper limit on the redshift will be presented in a forthcoming publication, where also a more detailed discussion of the constraints
on the source characteristics will be given.

\section*{Acknowledgements}

The support of the Namibian authorities and of the University of Namibia
in facilitating the construction and operation of H.E.S.S. is gratefully
acknowledged, as is the support by the German Ministry for Education and
Research (BMBF), the Max Planck Society, the French Ministry for Research,
the CNRS-IN2P3 and the Astroparticle Interdisciplinary Programme of the
CNRS, the U.K. Science and Technology Facilities Council (STFC),
the IPNP of the Charles University, the Polish Ministry of Science and 
Higher Education, the South African Department of
Science and Technology and National Research Foundation, and by the
University of Namibia. We appreciate the excellent work of the technical
support staff in Berlin, Durham, Hamburg, Heidelberg, Palaiseau, Paris,
Saclay, and in Namibia in the construction and operation of the
equipment.

This research made use of the NASA/IPAC Extragalactic Database (NED)  and of the SIMBAD Astronomical Database. The authors thank the {\it RXTE} team for their prompt response to our ToO request and the professional interactions that followed. The authors acknowledge the use of the publicly available {\it Swift} data, as well as the public HEASARC software packages. 

The authors wish to thank H.~Landt and M.~V\'eron-Cetty for very helpful discussions on the available redshift estimates of the source.

\end{document}